# Brain network similarity:
# Methods and applications

**Ahmad Mheich, Fabrice Wendling and Mahmoud Hassan**
LTSI, INSERM, U1099, Rennes, F-35000, France

**Abstract**

Graph theoretical approach has proved an effective tool to understand, characterize and quantify the complex brain network. However, much less attention has been paid to methods that quantitatively compare two graphs, a crucial issue in the context of brain networks. Comparing brain networks is indeed mandatory in several network neuroscience applications. Here, we discuss the current state of the art, challenges, and a collection of analysis tools that have been developed in recent years to compare brain networks. We first introduce the graph similarity problem in brain network application. We then describe the methodological background of the available metrics and algorithms of comparing graphs, their strengths and limitations. We also report results obtained in concrete applications from normal brain networks. More precisely, we show the potential use of brain network similarity to build a 'network of networks' that may give new insights into the object categorization in the human brain. Additionally, we discuss future directions in terms of network similarity methods and applications.

# I- INTRODUCTION

The human brain is a complex network that operates at multiple time/space scales. At the macro scale, the brain can be represented as a graph where the nodes denote the brain regions and the edges denote the connections (structural or functional) between these regions (Bullmore and Sporns 2009). The emerging field of Network Neuroscience has significantly improved our understanding about the structure and the function of the human brain (Bassett and Sporns 2017). Graph theory, a branch of mathematics focusing on understanding systems of interacting elements, has been shown as a very powerful tool to understand, characterize and quantify the complex brain network (Huang, Bolton et al. 2018, Yu, Du et al. 2018). Applying graph theoretical measures to brain network have revealed several non-trivial features such as small-worldness (Bassett and Bullmore 2017), modularity (Sporns and Betzel 2016) and a scale-free (van den Heuvel, Stam et al. 2008) behaviors. The usefulness of applying graph theory and network science to brain network analysis has been widely reviewed in the last decade from methodological (Wang and Chen 2003, Costa, Rodrigues et al. 2007) and applicative viewpoint (Christmas, Kittler et al. 1995, Luo and Hancock 2001, Cordella, Foggia et al. 2004, Bullmore and Sporns 2009, Hassan and Wendling 2018).

Surprisingly, much less attention has been paid to methods that quantitatively compare two graphs, a crucial issue in the context of brain networks. Comparing brain networks is indeed mandatory in several network neuroscience applications, including but not limited to, i) the estimation of the similarity between structural and functional brain networks, ii) the tracking of the temporal similarity of dynamic brain networks and iii) the computation of the (dis)similarity between normal and pathological brain networks or between two conditions (animals vs. tools) during a cognitive task (Sporns 2014, Liao, Vasilakos et al. 2017, Avena-Koenigsberger, Misic et al. 2018, Paban, Modolo et al. 2019, Rizkallah, Annen et al. 2019).

Quantifying similarity between networks is however complicated due the fact that complex networks, such as the brain, are composed of multi-scale (in time and space) systems whose structure and dynamics are difficult to encapsulate in a single score. The existing graph distance measures vary depending on the features used to compute the score: nodes, edges, spatial locations and spectrum (Christmas, Kittler et al. 1995, Luo and Hancock 2001, Wilson and Zhu 2008, Shimada, Hirata et al. 2016, Mheich, Hassan et al. 2018). Efforts to combine multiple features have been done very recently by several studies (Shimada, Hirata et al. 2016, Schieber, Carpi et al. 2017, Mheich, Hassan et al. 2018).

In this review, we will discuss the current state of the art, challenges, and a collection of analysis tools that have been developed in recent years to compare brain networks. We first introduce the graph similarity problem in brain network application. We then describe the methodological background of the available metrics and algorithms of comparing graphs, their strengths and limitations. From technical viewpoint, we describe two main families of methods: i) *graph theoretical approach* which consists of comparing the topological characteristics of two graphs at global, nodal or edge level (Bullmore and Sporns 2009, Zalesky, Fornito et al. 2010) and ii) *graph matching algorithms* including graphs and subgraphs isomorphism (Cordella, Foggia et al. 2004), graph edit distance (Gao, Xiao et al. 2010), kernel approach (Shervashidze, Vishwanathan et al. 2009) and other approaches (Cao, Li et al. 2013, Shimada, Hirata et al. 2016, Schieber, Carpi et al. 2017) .

From applicative viewpoint, we present new results using a recently developed algorithm called SimiNet (Mheich, Hassan et al. 2018), which takes into account the physical locations of nodes when computing similarity between two brain graphs. We show the potential use of network similarity in building a 'semantic map' of the brain (network of networks).

The paper is organized as follows: we first introduce the graph similarity problem in the context of network neuroscience. Second, we introduce the graph theoretical analysis and the methods and strategies used to compute distance between networks. Third, we present new results of the application for graph similarity methods in cognitive neuroscience.

## II- COMPARISON BETWEEN BRAIN NETWORKS

Brain graph model is an abstract mathematical representation of the interactions between brain elements (neurons, neural assemblies or brain regions). Nodes in this graph represent neuronal assemblies or brain regions obtained from certain parcellation techniques. Edges represent either functional (statistical dependence or level of synchronization between activity patterns of brain regions) or structural links (direct anatomical connections) between neural elements (Sporns 2010, Bullmore and Bassett 2011, Fornito, Zalesky et al. 2013, Fornito, Zalesky et al. 2016). Once the brain graph is constructed (and depending on the application), the comparison with other brain networks can be mandatory. Comparison techniques between brain networks have many applications due to the current widespread use of network neuroscience. These applications include statistical analysis between brain networks for different groups of subjects, or for the same subject before and after treatment, discrimination between neurological disorders by quantifying functional and topological similarities (Calderone, Formenti et al. 2016), quantifying the evolution of temporal brain networks at different time scales (Hassan, Benquet et al. 2015, Mheich, Hassan et al. 2015, O'Neill, Tewarie et al. 2017), comparing between real brain networks and generative network models (figure 1), and the comparison of the topological layout of nervous systems across species (Van den Heuvel, Bullmore et al. 2016).

| Notation | Description |
|---|---|
| $N, n$ | Set of nodes, number of nodes |
| $E, m$ | Set of edges, number of edges |
| $G$ | Graph |
| $\lambda$ | Eigenvalue |
| $C$ | Clustering coefficient |
| $L$ | Shortest path length |
| $S$ | Synchronizability |
| $BC$ | Betweenness centrality |
| $d$ | Density |
| $d_{Hamming}(G', G'')$ | Hamming distance between $G'$ and $G''$ |
| $d_{GED}(G', G'')$ | Graph edit distance between $G'$ and $G''$ |
| $A$ | Adjacency matrix |
| $\Lambda$ | Laplacian matrix |
| $D$ | Degree matrix |
| $k_i$ | Degree of node $i$ |

**Table 1: Notation and description**

Methods and strategies used to compare brain networks can be classified into two main classes (figure 2): the first one is the *statistical comparison*, where various graph theoretical metrics can be applied to characterize the topological architecture of the brain networks. There are metrics of global network organization and others can also be estimated at node or edge level of the compared networks (Bullmore and Bassett 2011). These metrics are then quantitatively compared between two groups of networks via statistical tests. This class includes also the *spectral analysis* with increased tendency to apply on brain networks. In the later, the comparison is based on the eigenvalues of adjacency/laplacian matrices of the compared graphs. The defined quantities and notations used in this paper are listed in table 1

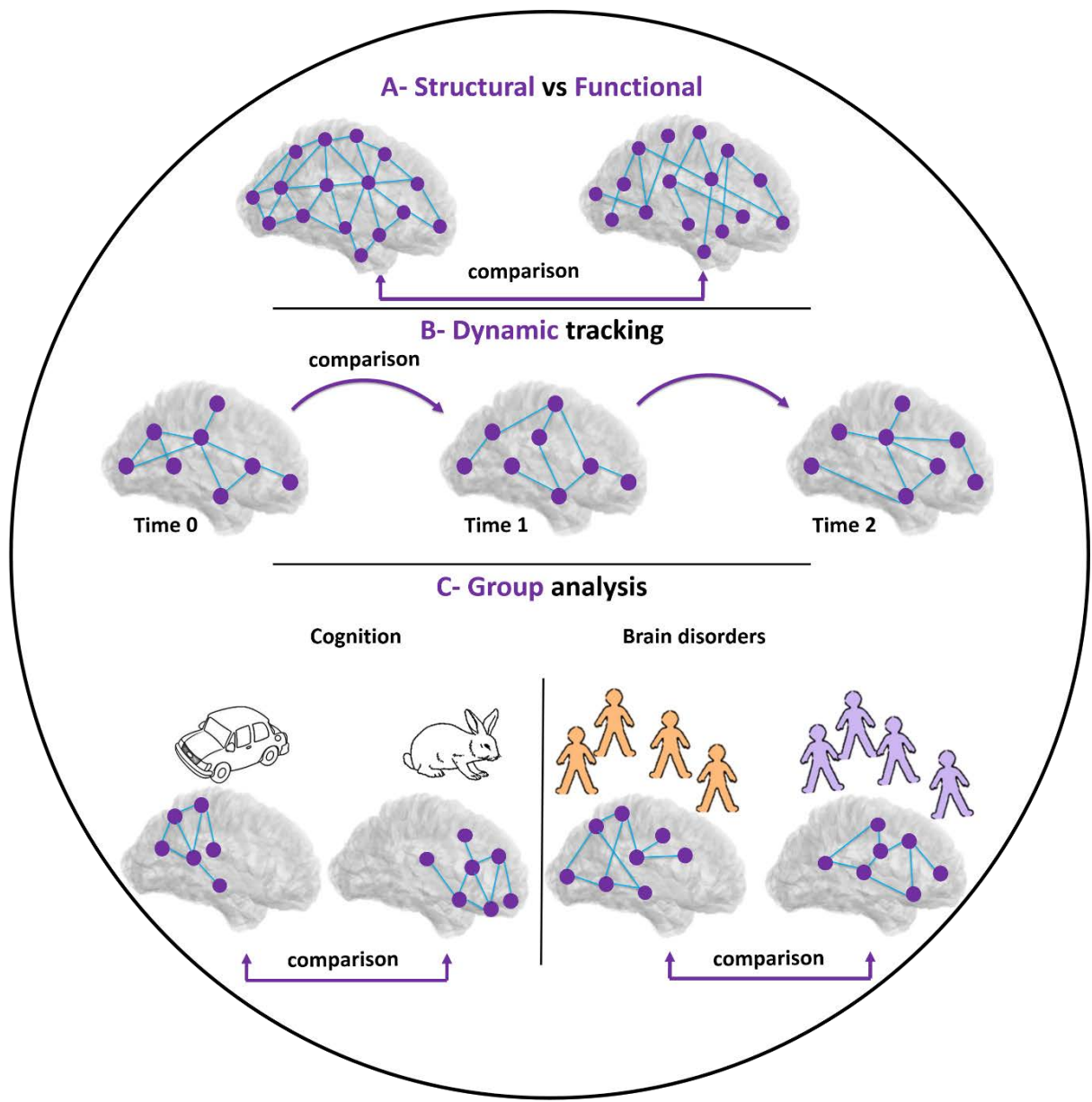

**Figure 1: Applications of graph comparison in network neuroscience. A- Comparison between structural and functional brain networks. B- Tracking the dynamic of brain networks during time. C- Comparison between two groups of brain networks for two different conditions.**

This type of comparison gives valuable information about the overall structure of a graph providing new insight into the workings of the brain as a complex system. The second class is the family of algorithms and strategies based on *graph matching*, where the main purpose is to quantify a distance (similarity score) between two networks by considering some characteristics that are supposed to be important from application viewpoint. Although most of the proposed algorithms are developed in specific domains, they represent promising tools to quantify similarity between brain networks.

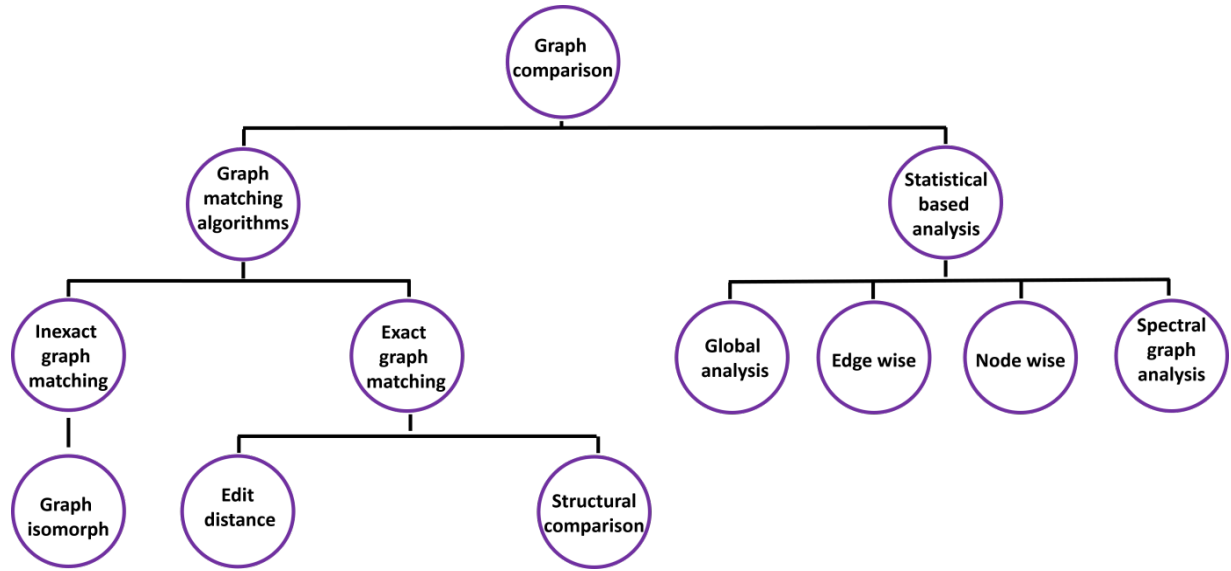

Figure 2: Graph comparison methods.

## 1. STATISTICAL COMPARISON

Statistical comparison between brains networks can be classified into two types: first, comparison of real brain networks to random networks, where the main purpose is to validate if some characteristics of the brain networks are significantly different than chance. For a given network, we can generate a large number of random graphs and by applying the graph theory metrics on these random graphs, as a point of reference, to test the randomness of the same metrics measured in the real brain networks. Second, the comparison of brain networks of two groups of subjects such as healthy control and patients via statistical hypotheses. We can classify metrics measured for comparing brain networks into four categories as following: global-level, node-wise, edge-wise and spectral graph analysis (figure 2). See (Bullmore and Sporns 2009, Zalesky, Fornito et al. 2010) for more details about the metrics. In the next paragraphs we will show a brief description about some selected metrics and their recent applications in brain network comparisons.

## 1.1 GLOBAL- LEVEL ANALYSIS

In this case, the graph metrics are computed over the entire network and one value can be derived per network. Statistical test are then applied to compare the two groups (such as healthy vs. patients).

### 1.1.1 Small worldness

Small worldness of a network was originally introduced by (Newman and Watts 1999). It is characterized by a low average shortest path length ($L$) and by high clustering coefficient ($C$). Briefly, the averaged path length $L$ is defined as the average minimum number of edges that have to be traversed to pass from one node to another in the network. The clustering coefficient ($C$) of a node is defined as the number of existing connections between the neighbors of the node divided by all the possible connections between them. $C$ quantifies the extent of local cliquishness or local efficiency of information transfer of a network.

It has been also reported that the functional brain networks derived from Alzheimer disease patients have the characteristics of random networks with characteristic path length significantly longer than the healthy subjects (Stam, De Haan et al. 2008). Other studies showed the presence of the small world characteristics in the brain connectivity of healthy subjects while these characteristics were disrupted in schizophrenia patients (Micheloyannis, Pachou et al. 2006, Liu, Liang et al. 2008, Lynall, Bassett et al. 2010).

### 1.1.2 Modularity

Network modules are defined by a subset of nodes in the graph that are densely connected to other nodes in the same module than the nodes in other modules (Girvan and Newman 2002, Sporns and Betzel 2016). The modularity consists of partitioning a network into a number of no-overlapping groups or modules-also called communities (Rubinov and Sporns 2010). Several methods have been proposed to resolve the community structure of complex networks in several applications (Newman 2006, Blondel, Guillaume et al. 2008, Lancichinetti and

Fortunato 2009). For brain networks applications, modularity maximization method (Newman and Girvan 2004) is the most applied in the detection of brain networks modules. The main idea of this method is to split the nodes of a network into $K$ non-overlapping communities in order to maximize the modularity quality function $Q$. A minimum value of $Q$ near to 0 indicates that the network considered is close to a random one, while a maximum value of $Q$ near to 1 indicates a strong community structure.

A number of studies found significant differences between the brain networks of two groups by comparing their modules. Meunier et al. (Meunier, Lambiotte et al. 2010) investigated the modular structure of the human brain networks derived from fMRI measurements for two groups of younger and older adults using modularity maximization algorithm. The results show that the brain becomes less modular with age, a finding reported as well by others (Baum, Ciric et al. 2017). Bassett et al. (Bassett, Wymbs et al. 2011) showed in a task-based fMRI study that brain modules become more segregated after training. In brain diseases applications, Alexander-Bloch et al. (Alexander-Bloch, Gogtay et al. 2010) showed that modularity decreases for schizophrenia patients compared to health subjects. In return, Peraza et al. (Peraza, Taylor et al. 2015) found an increase in modularity for patients with Lewy body diseases.

### 1.1.3 *Efficiency*

The network efficiency quantifies the exchange of information across the whole network and is defined as the inverse of the average path length (Latora and Marchiori 2001, Achard and Bullmore 2007). Global efficiency was used to compare the functional brain networks between two groups (healthy old subjects and healthy young subjects) (Achard and Bullmore 2007), authors showed that the efficiency was reduced in older people. In addition, many studies showed that patients with schizophrenia, Alzheimer and Parkinson diseases had a marked reduce in global efficiency compared with healthy controls (Skidmore, Korenkevych et al. 2011, Reijmer, Leemans et al. 2013, Berlot, Metzler-Baddeley et al. 2016).

### 1.2 NODE-WISE ANALYSIS

In this case, the graph metrics are calculated for each node and then the node's metric values are compared between the two graphs. The main advantages of such approach are i) the possibility to explore more features in the graph, ii) the presence of more data (number of nodes) to compare between conditions and iii) this comparison will not only show if there is a difference between two conditions but will also indicate where the difference is located (on which brain regions). However, it can produce 'false positive' results as the activity at each node is not fully independent. This type of analysis required correction for multiple comparisons, as comparison was done *n* times where *n* is the number of nodes, using methods such as Bonferroni (Rice 1989) or False Discovery Rate (*FDR*) (Genovese, Lazar et al. 2002). Several metrics can be computed at the level of network's nodes. For detailed and comprehensive review, authors can check Rubinov and Sporns (Rubinov and Sporns 2010). Globally speaking, these metrics reflect mainly three behaviors in the network:

#### *1.2.1  Segregation*

Segregation is the ability of the network for specialized processing in densely interconnected group of nodes. This includes measures such as i) Clustering coefficient (C) which is defined by the fraction of the nodes's neighbors that are also neighbors of each other (Watts and Strogatz 1998), and ii) Local efficiency which is a measure of the efficiency of information transfer limited to neighboring nodes. It is calculated as the average nodal efficiency among the neighboring nodes of node $i$, excluding node $i$ itself and iii) **Intra-module degree** which denotes how well connected is a node compared with other nodes of the same community. Chan et al. found a decrease in segregation of brain network for increasing age (Chan, Park et al. 2014). The network segregation was shown to be improved in Alzheimer and

Schizophrenia patients (He, Chen et al. 2008, Yao, Zhang et al. 2010, Zhang, Lin et al. 2012, Kabbara, Eid et al. 2018), and reduced for epilepsy patients (Zhang, Liao et al. 2011).

### *1.2.2 Integration*

Integration is the ability of the network to combine information from distributed nodes. This includes measures such as i) **participation coefficient** which quantifies the balance between the intra-module versus inter-module connectivity for a given node and ii) **characteristic path length** which is defined as the average shortest path length between all pairs of nodes in a network(Watts and Strogatz 1998). Several studies were performed to compare between brain networks of healthy subjects and patients diagnose with Alzheimer using characteristic path length (Stam, Jones et al. 2006, Supekar, Menon et al. 2008). The results showed an increase of characteristic path length in Alzheimer patients.

### *1.2.3 Hubness*

This include measures such as i) **Degree (strength)** that describes the connection strength of node to all other nodes, and ii) **Centrality** that describes the presence of nodes responsible for integrating total activity such as betweenness centrality. Betweenness centrality is defined as the fraction of all shortest paths in the network that pass through a given node. Many studies showed that the brain disorders are associated with alterations in the Hubs such as Alzheimer diseases, comatose patients and schizophrenia (Bassett, Bullmore et al. 2008, He, Chen et al. 2008, Lynall, Bassett et al. 2010, van den Heuvel, Mandl et al. 2010, Achard, Delon-Martin et al. 2012, Crossley, Mechelli et al. 2014). Yan et al. (Yan, Gong et al. 2010) used betweenness centrality to investigate the effects of sex on the topological organization of human cortical anatomical network. In clinical application, betweenness centrality was used to compare brain networks of healthy subjects and patients with schizophrenia, depression and Alzheimer diseases (van den Heuvel, Mandl et al. 2010, Yao, Zhang et al. 2010, Becerril, Repovs et al. 2011).

Others studies were performed to compare between brain networks of healthy subjects and patients diagnosed with schizophrenia using the degree of nodes (Bassett, Bullmore et al. 2008, Lynall, Bassett et al. 2010). The results showed a reduced degree in several brain nodes of schizophrenia patients. Other studies showed also that the Parkinson disease patients have a significant decrease in the degree of several brain regions in their functional network such as left dorsal lateral prefrontal cortex and had a significant increase in the degree of the left cerebellum (Wu, Wang et al. 2009).

## 1.3 EDGE-WISE ANALYSIS

The edge-wise analysis consists of calculating a statistical test (such as student *t-test*) on each edge in the graph. If the number of nodes in a graph equal to $n$ then the maximum number of edges (in the case of undirected network) is equal to $(n \times (n-1) / 2)$. The statistical test is calculated $(n \times (n-1) / 2)$ times. This method requires also correction for multiple comparisons using methods such as Bonferroni or *FDR*. Other approaches have been proposed also to deal with the family-wise error rate (*FWER*) sush as the network-based statistic (*NBS*) method (Zalesky, Fornito et al. 2010). The main idea of this method (based on permutation analysis) is to find a network "pattern" (a set of nodes connected by edges) instead of a single link that differentiates the two conditions. The *NBS* has been widely used to identify alterations in brain networks associated with psychiatric disorders such a schizophrenia and depression (Zalesky, Fornito et al. 2011), and to identify cognitive phenotypes in Parkinson's disease patients (Hassan, Chaton et al. 2017).

## 1.4 SPECTRAL GRAPH ANALYSIS

Spectral graph theory is a branch of graph theory which has been widely used to characterize the properties of a graph and extract information about its structure.

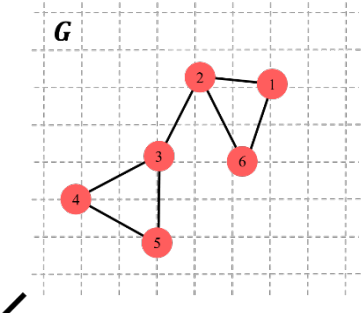
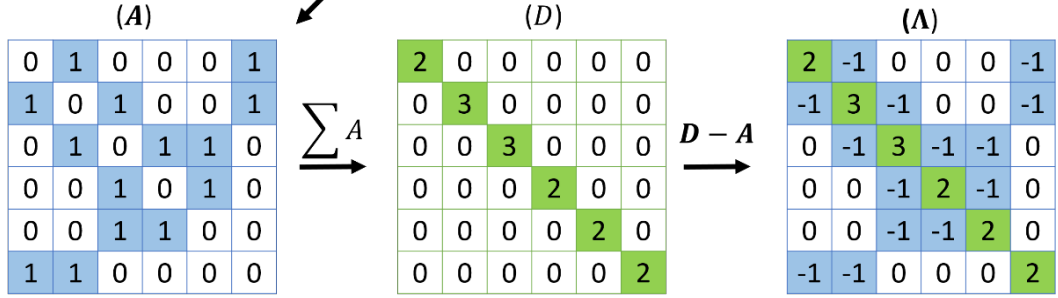
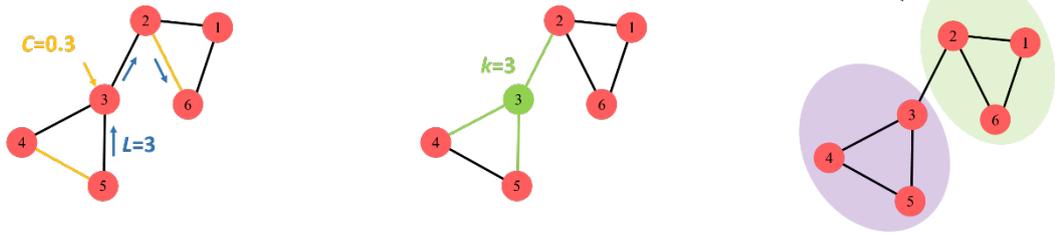

**Figure 3: a- A graph with six nodes and seven edges, b- (A) adjacency matrix, (D) degree matrix, (Λ) Laplacian matrix, c) Some of graph metrics extracted from each matrix.**

For a graph $G(N, E)$ of $n$ nodes with adjacency matrix $A_{n \times n}$ and degree matrix $D_{n \times n}$, the Laplacian matrix $\Lambda_{n \times n}$ is computed using the following formula (figure 3):

$$\Lambda(i,j) = \begin{cases} D_i & for\ i = j \\ -A(i,j) & for\ i \neq j \end{cases} where\ i, j \in N$$

Once the Laplacian matrix is constructed, the eigenvalue of $G$ can be computed ($\lambda_1, \lambda_2 ... \lambda_n$). Spectral graph analysis is well known in many domains for its powerful characterization of network properties (Farkas, Derényi et al. 2002, Banerjee 2012). It provides important information on relevant network properties such as connectivity levels and resilience to damage and the spread of information throughout the network (de Haan, van der Flier et al. 2012).

Comparing brain networks using the spectral graph theory was recently performed in several studies such as the comparison of network's eigenvalue distributions over the structural brain networks of different species such as caenorhabditis elegans, macaque, and cat (de Lange, de Reus et al. 2014). It was also used to detect network alterations in of Alzheimer disease patients (de Haan, van der Flier et al. 2012).

### *1.4.1 Synchronizability*

Synchronizability (*S*) quantifies the robustness of the network with respect to edge removals. It is computed as the report between the second smallest eigenvalue and the highest eigenvalue of the Laplacian matrix of the network.

$$S = \frac{\lambda_2}{\lambda_n}$$

A network with low value of *S* is more vulnerable for disconnection, in return a high *S* value means less vulnerable for disconnection. Several studies showed that many graph properties such as clustering coefficient, average distance, average degree and degree distribution failed to characterize the *S* of networks in return the spectral analysis can detect this *S* for the same networks (Atay, Bıyıkoğlu et al. 2006).

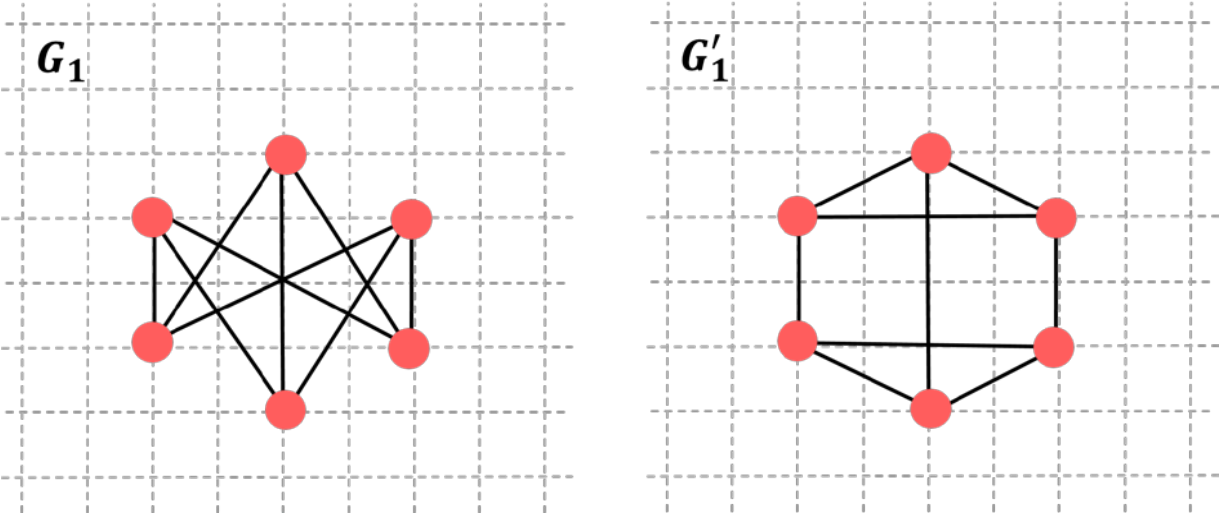

**Figure 4: Graph *G* and *G′***

For example, two graphs $G_1$ and $G_1'$ have the same number of nodes (*n*=6) and edges (*m*=9) shown in figure 4. These two graphs share common statistical network metrics such as: density, betweenness centrality, average degree and global efficiency (table 2), but they are differ in their *S*: $\lambda_2(G_1) = 3$ and $\lambda_2(G_1') = 2$, then $S(G_1)$=3/6 and $S(G_1')$=2/5.

| Graphs | Density | BC | Degree | Global Efficiency | S |
|---|---|---|---|---|---|
| $G_1$ | 0.6 | 2 | 3 | 0.8 | 0.5 |
| $G_1'$ | 0.6 | 2 | 3 | 0.8 | 0.4 |

**Table 2: Depiction of graph metrics: Density, BC (Betweennness centrality), degree, global efficiency and graph spectral metrics: S (synchronizability).**

De Haan et al. (de Haan, van der Flier et al. 2012) used graph spectral analysis to study synchronizability between healthy subjects and Alzheimer disease patients. Results showed a decrease in synchronizability and loss of network connectivity in most frequency bands for Alzheimer disease patients.

## 2 GRAPH MATCHING ALGORTIHMS

The general idea of graph matching consists of comparing two networks and providing a 'similarity' score. Methods of graph matching can be classified into two groups: exact graph matching and inexact graph matching. The result of comparing two graphs by an exact graph matching method (such as isomorphism-based measures (Cordella, Foggia et al. 2004)) is binary, in other words the compared graphs are either the same or not.

The inexact graph matching methods provide a similarity score between the compared graphs. This similarity value (if normalized) ranges between 0 (no similarity at all) to 1 (fully similar / same network). This includes the methods based on: (i) *edit distances* that focus on common and uncommon elements (nodes and edges) such as graph edit distance (GED) and hamming distance (Gao, Xiao et al. 2010) and (ii) *kernel methods* (Borgwardt and Kriegel 2005,

Shervashidze, Schweitzer et al. 2011) that focus on the *structure* of the networks by comparing their Laplacian matrices (figure 2).

## 2.1 ALGORITHMS BASED ON EDIT DISTANCE:

Quantifying the similarity/distance between two brain networks using edit distance algorithms allows to find the common/uncommon nodes (brain regions) and edges (functional /structural) between two brain networks.

*2.1.1* ***The hamming distance*** is the most directed way to compare two networks (Deza and Deza 2013). It is defined as the sum of difference between the adjacency matrices of two networks $G'$ and $G''$ :

$$d_{Hamming}(G', G'') = \sum_{i \neq j} A'_{ij} \neq A''_{ij}$$

Where $i$ and $j$ are two nodes and $A'$ and $A''$ the adjacency matrices of $G'$ and $G''$ respectively.

*2.1.2* ***The Graph Edit Distance (GED)*** is another popular distance between two networks (Gao, Xiao et al. 2010), widely applied in several applications (Zeng, Tung et al. 2009, Wang, Ding et al. 2012). It is defined as the minimum-weight sequence of edit operations required to transform one graph into another (an edit operation on a graph is an insertion, deletion or substitution applied on both nodes and edges). The graph edit distance between two graphs $G'$ and $G''$ is defined as:

$$d_{GED}(G', G'') = min \sum_{u \in U} c(e_u)$$

Where $c(e_u)$ is the cost of an edit operation from $G'$ to $G''$ and $U$ is the total number of edit operations. A challenging issue in this approach is to define the cost function for different operations.

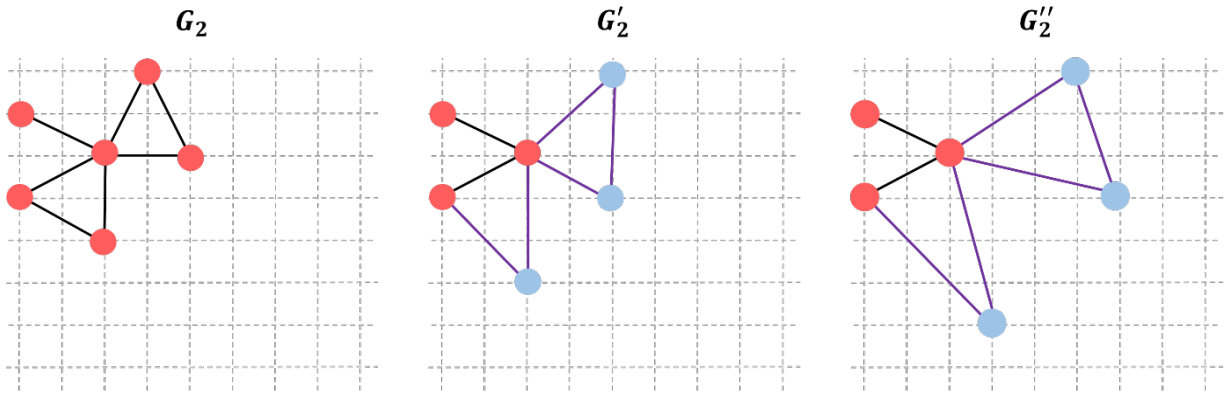

**Figure 5: Three networks with the same number of nodes (6) and edges (7) located into a grid (8 × 8).**

### 2.1.3 SimiNet algorithm

An important characteristic, not integrated in the previous approaches, is the spatial location of nodes, which denotes the 3D coordinates of the brain regions. The physical location of nodes can add additional key information when measuring similarity between brain networks. For instance, two networks with identical properties but interconnecting distant brain regions can have low similarity. Conversely, two networks with dissimilar properties but interconnecting spatially close brain regions can be very similar. SimiNet explore both the nodes and the edges when computing the similarity index. Concerning the nodes, the algorithm is based on four main steps: i) detection of common nodes between the two compared graphs, ii) substitution between two nodes where the cost of substitution equals to the distance between the substituted nodes, iii) insertion for new nodes where the cost of insertion equals to a constant value and iv) deletion of nodes where the cost of suppression equals to the cost of insertion. The (cost (substitution) < cost (insertion) + cost (deletion)) is always preserved. The second step is to calculate the edges distance. It consists on calculating the sum of the weight difference between two edges of two compared graphs. The algorithm provides a normalized Similarity Index (SI): 0 for no similarity and 1 for two identical networks (same properties and topology). The algorithm is detailed and compared with other methodologies in (Mheich, Hassan et al. 2018). Figure 5 displays three graphs $G_2$, $G'_2$ and $G''_2$ with the same number of nodes (*n*=6) and edges (*m*=7), these graphs are located into a grid (8 × 8). Graphs $G_2'$ and $G_2''$ are obtained by shifting three nodes of $G_2$ randomly. The similarity score is then calculated between the three graphs using Siminet, hamming and GED algorithms (table 3). As it can be seen in this example, Hamming and GED do not capture the spatial shifting of nodes, which is the case of SimiNet.

| Graphs | SimiNet | Hamming | GED |
| --- | --- | --- | --- |
| $(G_2, G_2')$ | 0.8 | 0.38 | 0.55 |
| $(G_2, G_2'')$ | 0.6 | 0.38 | 0.55 |
| $(G_2', G_2'')$ | 0.7 | 0.38 | 0.55 |

**Table 3: Depiction of similarity scores between the three networks ($G_2$, $G_2'$ and $G_2''$) using SimiNet, Hamming distance and graph edit distance (GED) algorithms.**

## 2.2 ALGORITHMS BASED ON STRUCTURE DISTANCE

Computing the similarity/distance between two brain networks using algorithms that prioritize network structures allow to spot and to quantify structural topology differences such as the presence or absence of important edges, nodes, cliques or sub-graphs that have influence on the information flow through the network. Several algorithms have been proposed to compute the network similarity based on structure distance. Some of these approaches and algorithms are briefly described hereafter:

### 2.2.1 Deltacon algorithm

The 'Deltacon' algorithm assesses the similarity for same size networks (two networks with same number of nodes) (Koutra, Vogelstein et al. 2013). The idea of this method is to compute the matrix of pairwise node affinities in the first network and to compare them with the one in the second network, where node affinities is the influence of each node on the other network's nodes. The difference between the matrices is then computed to produce an affinity score measuring the similarity between the compared networks. Readers may refer to (Koutra,

Vogelstein et al. 2013) for details about the Deltacon algorithm main steps and implementation. This algorithm also provides a normalized similarity ranging from 0 (dissimilar graphs) to 1 (identical graphs). Deltacon was applied to brain networks to classify people to groups of high and low creativity based on the similarity score between their brain connectivity graphs. DeltaCon satisfies some important properties from brain network applications viewpoint: i) **edge importance**, where edges that connect two components are of higher cost than other edges. ii) **Weight awareness** for weighted networks, where changes on edges with high weight values have more impact on the final similarity score and iii) **Edge sub-modularity**, where a specific change on network with few edges is more important than that in a denser network.

### 2.2.2 D-measure

Recently, Schieber et al. proposed a new algorithm to quantify graph dissimilarities by (Schieber, Carpi et al. 2017). The dissimilarity score is bordered between 0 and 1, where larger score correspond to more dissimilar graphs, and lower score to more similar graphs. The score produced by the algorithm is based on a combination of three components: i) dissimilarity in average node connectivity, ii) dissimilarity in a node dispersion metric, where node dispersion for a graph measures the distribution of distances between nodes in this graph and allows to make comparison with node dispersion in the second graph and iii) dissimilarity in node α-centrality (measure of nodes centrality within a graph). The main advantage of this algorithm is the ability to detect structural differences such as critical edges (connect two components) that have an influence on the information through the graph. D-measure was applied to brain networks in order to compare two groups of subjects (39 control and 68 alcoholic samples) (Schieber, Carpi et al. 2017). The algorithm was able to find the brain networks that discriminate control and alcoholic brain networks.

### 2.2.3 Kernel methods

Graph kernel methods are based on first mapping the graphs into a higher dimensional feature space and then searching for the common features among the mapping graphs. Given two graph $G_3$ and $G_3'$, the basic idea behind graph kernel is to construct a kernel $\xi(G_3, G_3',) = \langle \Phi(G_3), \Phi(G_3') \rangle$ where the similarity score between $G_3$ and $G_3'$ value corresponds to the scalar product between the two vector $\Phi(G_3)$ and $\Phi(G_3')$ in a Hilbert space. Several graph kernels-based algorithms have been proposed to measure network similarity such as random walks, shortest paths and Weisfeiler-Lehman.

- *Random walk* (Vishwanathan, Schraudolph et al. 2010): a random walk kernel counts the number of matching labeled random walks. The matching between two nodes is determined by comparing their attribute values. The measure of similarity between two random walks is then defined as the product of the kernel values corresponding to the nodes encountered along the walk.

- *Shortest path kernel* (Borgwardt and Kriegel 2005): computes the shortest path kernel for a set of graphs by exact matching of shortest path lengths. The Floyd-Warshall algorithm (Floyd 1962) is usually used to calculate all the pairs shortest-paths in $G_3$ and $G_3'$. The shortest path kernel is then defined by comparing all the pairs of the shortest path lengths among nodes in $G_3$ and $G_3'$.

- *Weisfeiler-Lehman* (Shervashidze, Schweitzer et al. 2011): computes h-step Weisfeiler-Lehman kernel for a set of graphs. The main idea of this algorithm is to increase the node labels by the sorted set of node labels of neighboring nodes, and compress these increased labels into new shorted labels. These steps are repeated until the node label sets of $G_3$ and $G_3'$, differ or the number of iterations reaches a maximum h. A detailed example is presented in figure 6.

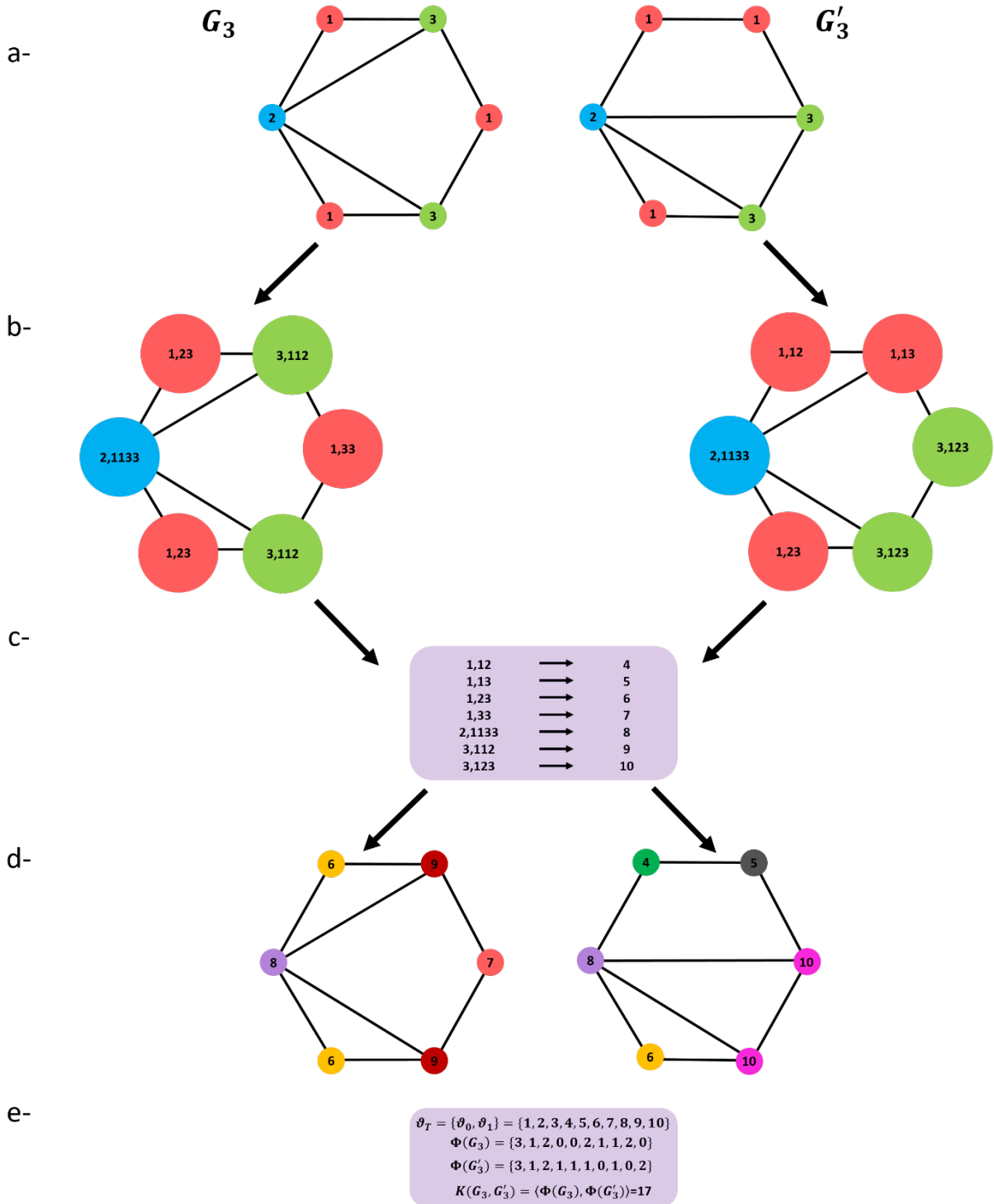

**Figure 6** Illustration of the construction process of the Weisfeiler-lehman subtree kernel with h=1 for two graphs $G_3$ and $G'_3$. a- the initial labeled graph $G_3$ and $G'_3$, b- augmented label on graph $G_3$ and $G'_3$, c- label compression, d- relabeled graph, e- computation of the kernel on graph $G_3$ and $G'_3$ where $\vartheta_0$ is the set of original node labels and $\vartheta_1$ the set of compressed node labels, $\Phi(G_3)$ and $\Phi(G'_3)$ are the count of node labels.

|  | | Characteristics | | | | |
|---|---|---|---|---|---|---|
| | | **Spatial location** | **Different size** | **Computational cost** | **Structural difference** | **Available Code** |
| **Methods** | **SimiNet** (Mheich, Hassan et al. 2018) | + | – | ++ | – | https://github.com/amheich/SimiNet |
| | **D-measure** (Schieber, Carpi et al. 2017) | – | – | + (not for sparse graph) | + | https://github.com/tischieber/Quantifying-Network-Structural-Dissimilarities |
| | **DeltaCon** (Koutra, Vogelstein et al. 2013) | – | – | + | + | http://web.eecs.umich.edu/~dkoutra/#tut |
| | **Kernel methods** (Borgwardt and Kriegel 2005, Shervashidze, Schweitzer et al. 2011) | – | + | – – | + | https://github.com/BorgwardtLab/GraphKernels |

**Table 4:** Advantages and limitations of graphs distance measures. Note that: '–' indicate a characteristic that not integrated in the similarity score of the method. '+' a characteristic that integrated in the methods. '– –' for worst computational time and '++' for very good computational time. 'Spatial location': physical location of nodes, 'Different size': Graphs with different number of nodes, 'Computational cost': algorithm running time, 'Structural difference': detection of difference between node's links in two graphs.

## III Network of networks

Analyzing similarity between brain networks can be useful for several applications in cognitive and clinical neuroscience. Here, we show an example of its application to functional networks estimated during visual object recognition task. To do so, we used dense-electroencephalography (256 electrodes) data from 20 subjects requested to name two categories of pictures (39 meaningful and 39 scrambled). Then, we construct a map based on the similarity scores between brain functional networks, figure 7.

This data is described in (Mheich, Hassan et al. 2018) and approved by the National Ethics Committee for the Protection of Persons (CPP), Braingraph study, agreement number (2014-A01461-46), and promoter: Rennes University Hospital.

Functional brain network for each object (picture) was constructed at the cortical level using the 'EEG-source connectivity' method (Hassan and Wendling 2018). The similarity scores between all the object-related functional networks were quantified using SimiNet algorithm, which produce a 78×78 similarity matrix. The similarity matrix was transformed into a graph where nodes represent brain networks and edges represent the highest similarity score between the brain networks. This graph is illustrated in figure 8. The visual inspection of this graph (blue nodes for meaningful and purple nodes for scrambled) shows that the connections between objects of the same category (N=72) is clearly higher than connections between objects from different categories (N=7). Constructing this '*network of networks*' can be seen as a first attempt to evaluate categorization of visual objects in the human brain from a functional network similarity-based approach.

## CONCLUSIONS

As long as there will be functional/structural brain networks, there will be people looking for comparing between them. Here, we have presented established views of the main methods and algorithms that can be used to compare brain networks.

Which method then? The answer depends on the application itself. If the objective is to reveal statistical difference between two groups (healthy vs. patients for instance) then the methods based in the graph theoretical approach (node-wide or edge-wise) can be good candidates provided that statistical parameters are carefully set (correction for multiple comparisons for instance). However, if the objective is to produce a similarity score (usually normalized between 0 and 1), then the graph matching methods are more appropriate.

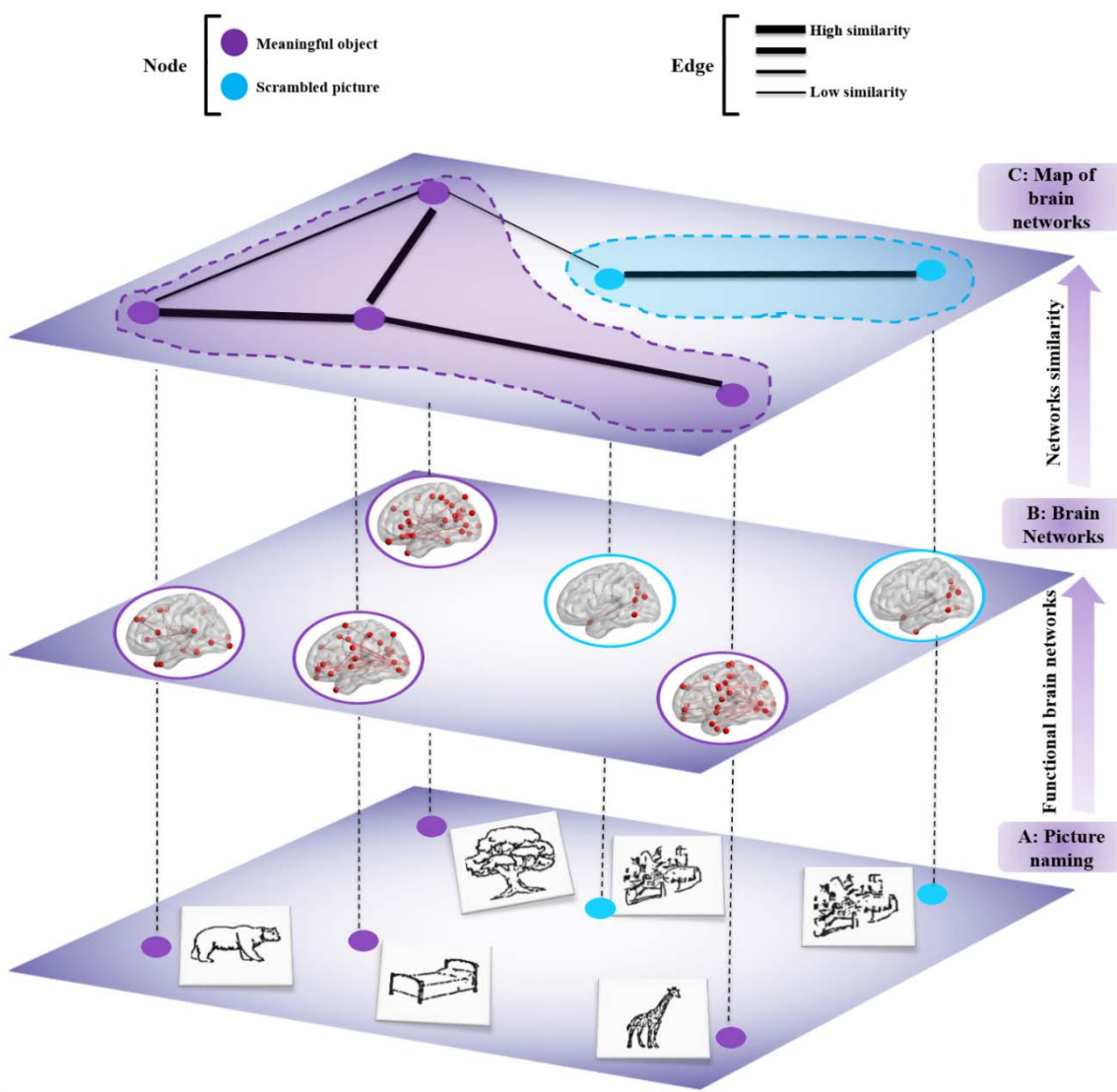

**Figure *7*: A schematic representation of the proposed method. A- Signal recording during meaningful and scrambled pictures naming task. B- Estimating the functional brain network for each picture. C- Measure the similarity between brain networks using SimiNet and classify them into meaningful and scrambled pictures.**

Validation of algorithms, like the comparative analysis of (Mheich et al. 2018), can allow to identify a set of methods that perform well on simulated networks. However, we do not know how well real networks are described by currently used simulations. Therefore, there is no guarantee that methods performing well on benchmarks also give reliable results on real brain networks (advantages and limitations of some of these algorithms are presented in table 4).

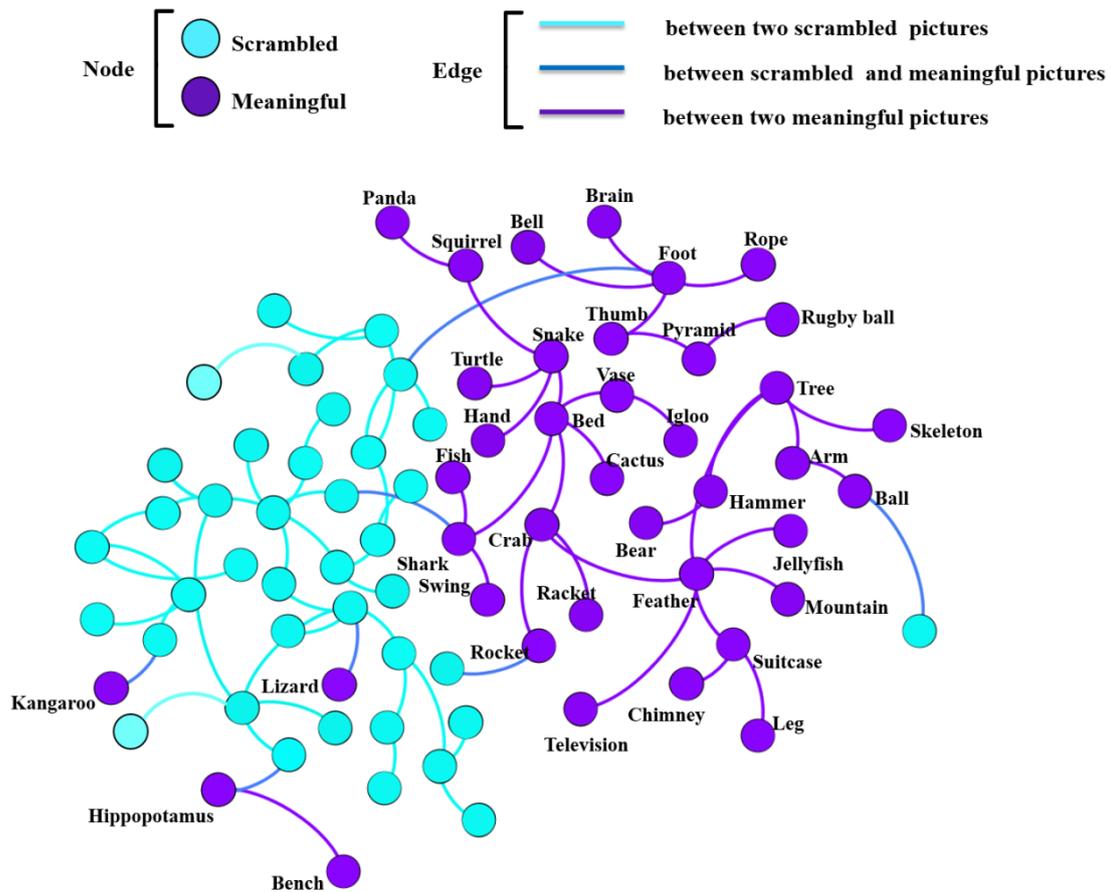

**Figure 8: Network of brain networks. The purple nodes represent meaningful objects and the blue nodes represent scrambled objects. The purple edges represent a high similarity value between two functional brain networks of meaningful category. The blue edges represent a high similarity value between two functional brain networks of scrambled category, and the dark blue edges represent a high similarity value between two brain networks of meaningful and scrambled objects.**

From application viewpoint, the network similarity distance is crucial in the identification of what we called here 'network of networks. This can used to build a 'semantic map' where nodes can represent the estimated networks of visual/auditory objects and edges can denote the similarity between these networks (preliminary results are presented in this review). This will undoubtedly require very large number of stimuli and also the repetition of each stimuli several times. When these conditions are respected, these semantic maps can give new insights into the object categorization process in the human brain from a network-based analysis.

In clinical neuroscience, a potential application of network distance measures is the mapping of a 'disease network' where the nodes may represent each brain disease and the edges can represent the similarity between the different networks associated to each disease (such as Parkinson's, Alzheimer's disease, epilepsy…). This application could help to further understand the possible common altered network patterns in brain disease. A very recent review by van den Heuvel and Sporns (van den Heuvel and Sporns 2019) showed indeed the importance of investigating such cross-disorder connectivity patterns.